\documentclass[a4paper,aps,onecolumn,nofootinbib]{revtex4}
\RequirePackage[colorlinks,hyperindex]{hyperref}
\RequirePackage[english]{babel}
\RequirePackage[latin1]{inputenc}
\RequirePackage[T1]{fontenc}
\RequirePackage{mathrsfs}
\RequirePackage{amsmath}
\RequirePackage{amssymb}
\RequirePackage{amsbsy}
\RequirePackage{color}
\RequirePackage{bm}
\hypersetup{colorlinks=true,breaklinks=true,urlcolor=blue,linkcolor=red}
\begin{document}
\title{\bf{Spinors beyond Dirac, Weyl and Majorana: the Flag-Dipoles}}
\author{Luca Fabbri$^{c}$\!\!\! $^{\hbar}$\!\!\! $^{G}$\footnote{luca.fabbri@unige.it}}
\affiliation{$^{c}$DIME, Universit\`{a} di Genova, Via all'Opera Pia 15, 16145 Genova, ITALY\\
$^{\hbar}$INFN, Sezione di Genova, Via Dodecaneso 33, 16146 Genova, ITALY\\
$^{G}$GNFM, Istituto Nazionale di Alta Matematica, P.le Aldo Moro 5, 00185 Roma, ITALY}
\date{\today}
\begin{abstract}
We recall the Lounesto classification of $1/2$-spin spinor fields, based on the vanishing of spinorial bilinear quantities: the classes are the regular spinor fields (i.e. the Dirac field), as well as singular spinor fields, also known as flag-dipole spinor fields, admitting two limiting sub-classes, given by the dipole spinors (i.e. the Weyl spinor) and the flagpole spinors (i.e. the Majorana spinor). We discuss each class in terms of its representatives, with particular emphasis upon the flag-dipole spinor fields.
\end{abstract}
\maketitle
\section{Introduction}
It is common knowledge of present-day field theory that all spinor fields are either Dirac, or Weyl or Majorana.

Yet, from a purely mathematical perspective, these three spinor fields do not appear to extinguish all possibilities: take for instance the Lounesto classification, which is based on the vanishing of spinorial bilinear quantities.

For the $1/2$-spin spinor fields, these spinorial bilinear quantities are the scalar $\Phi$ and the pseudo-scalar $\Theta$, the vector $U_{i}$ and the axial-vector $S_{i}$, and the antisymmetric tensor $M_{ab}$ (their definition will be given below). Since $U_{i}\!\neq\!0$ and in view of the Fierz identities, there are only four ways in which the bilinear quantities can vanish. The first splitting comes from having either at least one of the two scalars not identically zero, giving a class called \emph{regular}, or $\Phi\!=\!\Theta\!=\!0$, giving a class called \emph{singular}. The elements of the singular class are also known as \emph{flag-dipoles}, and they contain two sub-classes having either $M_{ab}\!=\!0$, whose elements are called \emph{dipoles}, or $S_{i}\!=\!0$, whose elements are called \emph{flagpoles} \cite{L}.

Regular, flag-dipoles in proper sense, dipoles and flagpoles are therefore the four exhaustive classes of the Lounesto classification.\footnote{Actually, the original Lounesto classification in \cite{L} consisted of six classes, since the regular class is split into three sub-classes. However, such a distinction will be irrelevant for us, and consequently we will make no further splitting of the regular class.} Now, it is known that the regular class has elements recognized to be the Dirac spinor fields, much in the same way we know that dipoles and flagpoles are respectively the Weyl and Majorana spinors \cite{Cavalcanti:2014wia,HoffdaSilva:2017waf,daSilva:2012wp,R1,R2,Rogerio:2020ewe,Cavalcanti:2020obq,Ablamowicz:2014rpa,daRocha:2008we,BuenoRogerio:2019tvv}. Consequently, it is an unclear circumstance that the flag-dipoles in the proper sense seem to have no counterpart in field theory.

The reticence of field theorists to consider flag-dipoles may be due to the fact that, at a classical level, there does not appear to be any spinor with the structure of flag-dipole and that is also a solution to the spinor equation. In what can be found in literature, there are only two examples, and both in very specific non-trivial backgrounds \cite{daRocha:2013qhu,Meert:2018qzk}.

In the present paper we provide another example. But more than that, our example of flag-dipole solving the spinor equation will be found with a method that is general enough to open the way to more solutions of the same type.
\section{Geometric Classification: Polar Form}
We begin by setting the notation and conventions, starting from the Clifford matrices $\boldsymbol{\gamma}_{a}$ ($\boldsymbol{\gamma}$ standing for ``Clifford'') defined to verify $\left\{\boldsymbol{\gamma}_{a},\boldsymbol{\gamma}_{b}\right\}\!=\!2\eta_{ab}\mathbb{I}$ where $\eta_{ab}$ is the Minkowskian metric: the sigma matrices $\boldsymbol{\sigma}_{ab}$ ($\boldsymbol{\sigma}$ standing for ``spin'') defined as $\boldsymbol{\sigma}_{ab}\!=\!\left[\boldsymbol{\gamma}_{a},\boldsymbol{\gamma}_{b}\right]/4$ are generators of spinor transformations. The parity-odd matrix $\boldsymbol{\pi}$ ($\boldsymbol{\pi}$ standing for ``parity'') is implicitly defined by $2i\boldsymbol{\sigma}_{ab}\!=\!\varepsilon_{abcd}\boldsymbol{\pi}\boldsymbol{\sigma}^{cd}$ where $\varepsilon_{abcd}$ is the completely antisymmetric pseudo-tensor of Levi-Civita.

To explicitly write these matrices, when needed, we will employ the chiral representation.
\subsection{Kinematics: algebra}
A spinor $\psi$ is a column of $4$ complex fields, each of which being a scalar for diffeomorphisms, and such that altogether transform as $\psi\!\rightarrow\!\boldsymbol{S}\psi$ where $\boldsymbol{S}\!=\!\exp{(iq\alpha)}\boldsymbol{\Lambda}$: in this expression $\exp{(iq\alpha)}$ represents the abelian $U(1)$ gauge of charge $q$ and parameter $\alpha$, while $\boldsymbol{\Lambda}\!=\!\exp{(\theta^{ab}\boldsymbol{\sigma}_{ab}/2)}$ is an element of the complex Lorentz group with parameters $\theta^{ab}\!=\!-\theta^{ba}$ (that is, belonging to the spin-$1/2$ representation of the Lorentz group). The adjoint spinor is defined by $\overline{\psi}\!=\!\psi^{\dagger}\boldsymbol{\gamma}^{0}$ as the only dual transforming like $\overline{\psi}\!\rightarrow\!\overline{\psi}\boldsymbol{S}^{-1}$ while preserving parity and built exclusively with Clifford matrices. Given the transformation properties of the Clifford matrices, the spinor bilinear quantities
\begin{gather}
2\overline{\psi}\boldsymbol{\sigma}^{ab}\boldsymbol{\pi}\psi\!=\!\Sigma^{ab}\ \ \ \ \ \ \ \ \ \ \ \
2i\overline{\psi}\boldsymbol{\sigma}^{ab}\psi\!=\!M^{ab}\label{tensors}\\
\overline{\psi}\boldsymbol{\gamma}^{a}\boldsymbol{\pi}\psi\!=\!S^{a}\ \ \ \ \ \ \ \ \ \ \ \
\overline{\psi}\boldsymbol{\gamma}^{a}\psi\!=\!U^{a}\\
i\overline{\psi}\boldsymbol{\pi}\psi\!=\!\Theta\ \ \ \ \ \ \ \ \ \ \ \
\overline{\psi}\psi\!=\!\Phi
\end{gather}
are all real tensors (in the left column the parity-odd ones, in the right column the parity-even ones). Straightforwardly, we have $\Sigma^{ab}\!=\!-\varepsilon^{abij}M_{ij}/2$ (showing that the two tensors in (\ref{tensors}) are the Hodge dual of one another) and also
\begin{gather}
M_{ab}\Theta\!+\!\Sigma_{ab}\Phi\!=\!U_{[a}S_{b]}\ \ \ \ \ \ \ \
\ \ \ \ \ \ \ \ M_{ab}\Phi\!-\!\Sigma_{ab}\Theta\!=\!U^{j}S^{k}\varepsilon_{jkab}\label{A}
\end{gather}
together with
\begin{gather}
\Sigma_{ik}U^{i}\!=\!\Phi S_{k}\ \ \ \ \ \ \ \
\ \ \ \ \ \ \ \ M_{ik}U^{i}=\Theta S_{k}\\
M_{ik}S^{i}=\Theta U_{k}\ \ \ \ \ \ \ \
\ \ \ \ \ \ \ \ \Sigma_{ik}S^{i}\!=\!\Phi U_{k}
\end{gather}
and
\begin{gather}
\frac{1}{2}M_{ab}M^{ab}\!=\!-\frac{1}{2}\Sigma_{ab}\Sigma^{ab}\!=\!\Phi^{2}\!-\!\Theta^{2}\label{norm1}\\
\frac{1}{2}M_{ab}\Sigma^{ab}\!=\!-2\Theta\Phi
\label{orthogonal1}
\end{gather}
\begin{gather}
U_{a}U^{a}\!=\!-S_{a}S^{a}\!=\!\Theta^{2}\!+\!\Phi^{2}\label{norm2}\\
U_{a}S^{a}\!=\!0\label{orthogonal2}
\end{gather}
known in literature as Fierz re-arrangements, all of which being geometric identities (we will not give the demonstration for these identities since they constitute a very basic list of identities of any course on spinors --- we leave to interested readers the task of checking their form against those that are presented in reference \cite{daSilva:2012wp} or reference \cite{Ablamowicz:2014rpa}).

With these identities, it is now possible to demonstrate the following theorem \cite{jl, Fabbri:2016msm}. If a spinor is such that at least one between $\Theta$ or $\Phi$ is not identically zero, it is called \emph{regular} spinor and it is possible to write it in polar form
\begin{gather}
\psi\!=\!\phi e^{-\frac{i}{2}\beta\boldsymbol{\pi}}
\boldsymbol{L}^{-1}\left(\!\begin{tabular}{c}
$1$\\
$0$\\
$1$\\
$0$
\end{tabular}\!\right)
\label{regular}
\end{gather}
where
\begin{gather}
\Theta\!=\!2\phi^{2}\sin{\beta}\ \ \ \ \ \ \ \ \ \ \ \ \ \ \ \  \Phi\!=\!2\phi^{2}\cos{\beta}\\
S^{a}\!=\!2\phi^{2}s^{a}\ \ \ \ \ \ \ \ \ \ \ \ \ \ \ \  U^{a}\!=\!2\phi^{2}u^{a}
\end{gather}
verifying $u_{a}u^{a}\!=\!-s_{a}s^{a}\!=\!1$ and $u_{a}s^{a}\!=\!0$ and in terms of which
\begin{gather}
M^{ab}\!=\!2\phi^{2}(\cos{\beta}u_{j}s_{k}\varepsilon^{jkab}\!+\!\sin{\beta}u^{[a}s^{b]})
\end{gather}
showing that the angular momentum $M^{ab}$ is written in terms of the velocity $u^{a}$ and the spin $s^{a}$ as well as $\phi$ (called module) and $\beta$ (called chiral angle). If $\Theta\!=\!\Phi\!\equiv\!0$, it is said \emph{singular} spinor and it is possible to write it in polar form
\begin{gather}
\psi\!=\!\frac{1}{\sqrt{2}}\left(\mathbb{I}\cos{\frac{\alpha}{2}}
\!-\!\boldsymbol{\pi}\sin{\frac{\alpha}{2}}\right)\boldsymbol{L}^{-1}\left(\!\begin{tabular}{c}
$1$\\
$0$\\
$0$\\
$1$
\end{tabular}\!\right)
\label{singular}
\end{gather}
so that $U_{a}U^{a}\!=\!0$ as well as $M_{ab}M^{ab}\!=\!\varepsilon^{abij}M_{ab}M_{ij}\!=\!0$ and $M_{ik}U^{i}\!=\!\varepsilon_{ikab}M^{ab}U^{i}\!=\!0$ and where
\begin{gather}
S^{a}\!=\!-\sin{\alpha}U^{a}\label{spin}
\end{gather}
showing that the spin is written in terms of the velocity as well as $\alpha$ (called flag-dipole angle). For both, $\boldsymbol{L}$ has the structure of spinor transformations, the transformations ensuring that the polar form preserve manifest covariance.

Regular spinor fields are the Dirac spinor fields \cite{Cavalcanti:2014wia,HoffdaSilva:2017waf,daSilva:2012wp,R1,R2,Rogerio:2020ewe,Cavalcanti:2020obq,Ablamowicz:2014rpa,daRocha:2008we,BuenoRogerio:2019tvv}. Singular spinor fields admit a further sub-classification \cite{Fabbri:2020elt}: if no constraint is made on $\alpha$, the spinor can also be called \emph{flag-dipole} and it is generally given by (\ref{singular}). If $\alpha\!=\!\pm\pi/2$, the spinor is called \emph{dipole} and is given by either
\begin{gather}
\psi\!=\!\boldsymbol{L}^{-1}\left(\!\begin{tabular}{c}
$1$\\
$0$\\
$0$\\
$0$
\end{tabular}\!\right)\ \ \ \ \ \ \ \ \ \ \ \ \ \ \ \ \mathrm{or}\ \ \ \ \ \ \ \ \ \ \ \ \ \ \ \
\psi\!=\!\boldsymbol{L}^{-1}\left(\!\begin{tabular}{c}
$0$\\
$0$\\
$0$\\
$1$
\end{tabular}\!\right)
\label{dipole}
\end{gather}
having $S^{a}\!=\!-U^{a}$ or $S^{a}\!=\!U^{a}$ and in both cases
\begin{gather}
M^{ab}\!=\!0
\end{gather}
showing the absence of the angular momentum. If $\alpha\!=\!0$, the spinor is called \emph{flagpole} and is given by
\begin{gather}
\psi\!=\!\boldsymbol{L}^{-1}\left(\!\begin{tabular}{c}
$1$\\
$0$\\
$0$\\
$1$
\end{tabular}\!\right)
\label{flagpole}
\end{gather}
having
\begin{gather}
S^{a}\!=\!0
\end{gather}
showing the absence of the spin. As above, in all cases, $\boldsymbol{L}$ has the structure of a spinorial transformation.

Flagpoles are known in physics as Majorana spinors (that is eigen-spinors of charge conjugation), while dipoles are known in physics as Weyl spinors (that is eigen-spinors of chirality) \cite{Fabbri:2020elt}. As $U^{a}\!\neq\!0$ such classification is exhausted.

It is important to stress that every one of the above representations of spinor fields (\ref{regular}), (\ref{singular}), (\ref{dipole}) and (\ref{flagpole}) is always defined up to some helicity flip and up to the $\psi\!\rightarrow\!\boldsymbol{\pi}\psi$ discrete transformation.

Therefore the total number of mutually-exclusive and exhaustive types amounts to four.\footnote{In making a parallel to the Lounesto classification, regular spinors are in classes I, II and III, singular spinors in the form of flag-dipoles are in class IV, while dipoles and flagpoles are in classes VI and V \cite{L}.}
\subsection{Kinematics: derivatives}
In its essence, this classification is purely algebraic. Nevertheless, we will now assess the differential features of all four classes. For regular spinor fields it is known since quite some time now that it is always possible to find one real vector $P_{\mu}$ (called linear momentum) and a real antisymmetric tensor $R_{ab\mu}\!=\!-R_{ba\mu}$ (called tensorial connection) \cite{Fabbri:2018crr} with which we can write in general
\begin{gather}
\boldsymbol{\nabla}_{\mu}\psi\!=\!\left(\nabla_{\mu}\ln{\phi}\mathbb{I}
\!-\!\frac{i}{2}\nabla_{\mu}\beta\boldsymbol{\pi}
\!-\!iP_{\mu}\mathbb{I}\!-\!\frac{1}{2}R_{ij\mu}\boldsymbol{\sigma}^{ij}\right)\psi
\label{decspinder-reg}
\end{gather}
as the polar form of the spinor field's covariant derivative \cite{Fabbri:2024avj}. We also have
\begin{gather}
\nabla_{\mu}s_{i}\!=\!R_{ji\mu}s^{j}\ \ \ \ \ \ \ \
\ \ \ \ \ \ \ \ \nabla_{\mu}u_{i}\!=\!R_{ji\mu}u^{j}\label{ds-du}
\end{gather}
as geometric identities. Similarly, for singular spinor fields the same $P_{\mu}$ and $R_{ab\mu}$ allow us to write
\begin{gather}
\boldsymbol{\nabla}_{\mu}\psi\!=\!\left[-\frac{1}{2}\left(\mathbb{I}\tan{\alpha}
\!+\!\boldsymbol{\pi}\sec{\alpha}\right)\nabla_{\mu}\alpha
\!-\!iP_{\mu}\mathbb{I}\!-\!\frac{1}{2}R_{ij\mu}\boldsymbol{\sigma}^{ij}\right]\psi
\label{decspinder-sing}
\end{gather}
as the polar form of the spinor field's covariant derivative \cite{Fabbri:2020elt}. Again
\begin{gather}
\nabla_{\mu}M^{ab}\!=\!-M^{ab}\tan{\alpha}\nabla_{\mu}\alpha
\!-\!R^{a}_{\phantom{a}k\mu}M^{kb}\!+\!R^{b}_{\phantom{b}k\mu}M^{ka}\label{dM}\\
\nabla_{\mu}U_{i}\!=\!R_{ji\mu}U^{j}\label{dU}
\end{gather}
are valid as geometric identities. The two limiting cases of dipoles and flagpoles have a constant $\alpha$ so that it is clear to see that for dipoles we have
\begin{gather}
\boldsymbol{\nabla}_{\mu}\psi
\!=\!\left(-iP_{\mu}\mathbb{I}\!-\!\frac{1}{2}R_{ij\mu}\boldsymbol{\sigma}^{ij}\right)\psi
\end{gather}
in general. The same is true also for flagpoles although now we have
\begin{gather}
\boldsymbol{\nabla}_{\mu}\psi\!=\!-\frac{1}{2}R_{ij\mu}\boldsymbol{\sigma}^{ij}\psi
\end{gather}
since in this case the linear momentum also vanishes.

We notice that, for all four forms of covariant derivatives, there are terms proportional to the $\mathbb{I}$, $\boldsymbol{\pi}$ and $\boldsymbol{\sigma}^{ij}$ matrices, but never any term proportional to Clifford matrices: this is because the $\boldsymbol{\gamma}^{k}$ matrices are off-diagonal and hence their presence would mix chiralities. Notice also that for flagpoles the linear momentum vanishes due to charge-conjugation: interested readers can have a look at reference \cite{Fabbri:2024eec} for a detailed discussion on the subject.
\subsection{Dynamics}
The dynamical character of $1/2$-spin spinor fields are assigned by the Dirac differential equation
\begin{gather}
i\boldsymbol{\gamma}^{\mu}\boldsymbol{\nabla}_{\mu}\psi\!-\!m\psi\!=\!0
\label{D}
\end{gather}
where $m$ is the mass. Expressing the covariant derivative of spinors in polar form, (\ref{D}) can be converted in polar form as well. For this purpose, we set
\begin{gather}
\frac{1}{2}\varepsilon_{\mu\alpha\nu\iota}R^{\alpha\nu\iota}\!=\!B_{\mu}\\
R_{\mu a}^{\phantom{\mu a}a}\!=\!R_{\mu}
\end{gather}
to keep the notation compact. Then, for regular spinor fields, the Dirac equation is equivalent to the pair
\begin{gather}
B_{\mu}\!-\!2P^{\iota}u_{[\iota}s_{\mu]}
\!+\!\nabla_{\mu}\beta\!+\!2s_{\mu}m\cos{\beta}\!=\!0\label{dep1}\\
R_{\mu}\!-\!2P^{\rho}u^{\nu}s^{\alpha}\varepsilon_{\mu\rho\nu\alpha}
\!+\!2s_{\mu}m\sin{\beta}\!+\!\nabla_{\mu}\ln{\phi^{2}}\!=\!0\label{dep2}
\end{gather}
as demonstrated and thoroughly discussed in \cite{Fabbri:2025ftm}. For singular spinor fields, the Dirac equation is equivalent to
\begin{gather}
(\varepsilon^{\mu\rho\sigma\nu}\nabla_{\mu}\alpha\sec{\alpha}
\!-\!2P^{[\rho}g^{\sigma]\nu})M_{\rho\sigma}\!=\!0
\label{fd1}\\
M_{\rho\sigma}(g^{\nu[\rho}\nabla^{\sigma]}\alpha\sec{\alpha}
\!-\!2P_{\mu}\varepsilon^{\mu\rho\sigma\nu})\!+\!4m\sin{\alpha}U^{\nu}\!=\!0
\label{fd2}\\
[-B^{\sigma}\varepsilon_{\sigma\mu\rho\nu}\!+\!R_{[\mu}g_{\rho]\nu}
\!+\!g_{\nu[\mu}\nabla_{\rho]}\alpha\tan{\alpha}]M_{\eta\zeta}\varepsilon^{\mu\rho\eta\zeta}\!=\!0
\label{fd3}\\
[-B^{\sigma}\varepsilon_{\sigma\mu\rho\nu}\!+\!R_{[\mu}g_{\rho]\nu}
\!+\!g_{\nu[\mu}\nabla_{\rho]}\alpha\tan{\alpha}]M^{\mu\rho}\!+\!4mU_{\nu}\!=\!0
\label{fd4}
\end{gather}
as discussed in \cite{Fabbri:2020elt}. The limiting case of dipole spinors has Dirac equation reducing to
\begin{gather}
R_{\mu}U^{\mu}\!=\!0\label{d1}\\
(-B_{\mu}\!\pm\!2P_{\mu})U^{\mu}\!=\!0\label{d2}\\
[(-B_{\mu}\!\pm\!2P_{\mu})\varepsilon^{\mu\rho\alpha\nu}
\!+\!g^{\rho[\alpha}R^{\nu]}]U_{\rho}\!=\!0\label{d3}
\end{gather}
with constraint $m\!=\!0$ automatically fixed \cite{Fabbri:2021mfc}. The limiting case of flagpole spinors has Dirac equation reducing to
\begin{gather}
R_{\mu}U^{\mu}\!=\!0\label{f1}\\
B_{\mu}U^{\mu}\!=\!0\label{f2}\\
(-B_{\mu}\varepsilon^{\mu\rho\alpha\nu}
\!+\!g^{\rho[\alpha}R^{\nu]})U_{\rho}\!+\!2m\!M^{\alpha\nu}\!\!=\!0\label{f3}
\end{gather}
or equivalently
\begin{gather}
(g_{\sigma[\pi}B_{\kappa]}\!-\!R^{\mu}\varepsilon_{\mu\sigma\pi\kappa})M^{\pi\kappa}\!=\!0\label{f5}\\
\frac{1}{2}(B_{\mu}\varepsilon^{\mu\sigma\pi\kappa}
\!+\!g^{\sigma[\pi}R^{\kappa]})M_{\pi\kappa}\!-\!2mU^{\sigma}\!=\!0\label{f6}
\end{gather}
as shown in reference \cite{Fabbri:2021mfc}.

See also \cite{Fabbri:2024eec} for a general overview. It is to be noticed that the Dirac equations are as many as their corresponding polar form for the regular spinor fields, singular spinor fields have their polar equations that amount to twice as many, therefore displaying a two-fold redundancy \cite{Fabbri:2020elt}. This redundancy has not been discussed in any detail, yet.

In the next section we will focus on flag-dipole spinor fields, trying to clarify this issue.
\section{Flag-Dipole Spinor Fields}
\subsection{Structure of the field equations}
Until now, we reviewed the general theory of spinor fields for all four types in polar formulation \cite{Fabbri:2024eec, Fabbri:2020elt}.

From this section on we will develop new material, focusing on flag-dipole spinor fields, with the goal of clarifying the results of \cite{Fabbri:2020elt}, where the flag-dipole spinor field equations were presented, but their redundancy not discussed.

The fundamental step in this direction is to notice that, for flag-dipoles, there always exists a vector $X_{a}$ such that $X_{a}X^{a}\!=\!-1$, as well as $X_{a}U^{a}\!=\!0$, with which
\begin{gather}
M^{ab}\!=\!\cos{\alpha}U^{[a}X^{b]}\label{M}
\end{gather}
as was demonstrated in Appendix \ref{app-M}. Consequently, we have
\begin{gather}
U^{a}\!=\!-\sec{\alpha}M^{ab}X_{b}\label{U}
\end{gather}
as easy to prove with a straightforward inversion. We also have that
\begin{gather}
\Sigma^{ab}\!=\!-\cos{\alpha}U_{i}X_{j}\varepsilon^{ijab}\label{Sigma}
\end{gather}
which is equivalent to (\ref{M}) since they are the Hodge dual of one another. Eventually
\begin{gather}
\Sigma^{ab}X_{b}\!=\!0\label{Zero}
\end{gather}
as is obvious to verify.

Because of $U_{a}U^{a}\!=\!0$ the check of $M_{ik}U^{i}\!=\!\varepsilon_{ikab}M^{ab}U^{i}\!=\!0$ as well as $M_{ab}M^{ab}\!=\!\varepsilon^{abij}M_{ab}M_{ij}\!=\!0$ is immediate.

We will finally assume that the vector $X_{a}$ verify $\nabla_{\mu}X_{a}\!=\!R_{ka\mu}X^{k}$ identically.

It is then possible to see that, because of (\ref{dU}), expression (\ref{dM}) is satisfied, as shown in Appendix \ref{app-der}.

At last, we have that the flag-dipole field equations (\ref{fd1}-\ref{fd4}) can be reduced to the field equations
\begin{gather}
(2\sin{\alpha}P_{\pi}\!-\!B_{\pi})U_{\mu}\varepsilon^{\pi\mu\rho\sigma}
\!-\!R^{[\rho}U^{\sigma]}
\!+\!2mM^{\rho\sigma}\!=\!0\label{fd1red}\\
U_{[\rho}\nabla_{\sigma]}\alpha
\!+\!2\cos{\alpha}P^{\pi}U^{\mu}\varepsilon_{\pi\mu\rho\sigma}
\!-\!2m\tan{\alpha}M_{\rho\sigma}\!=\!0\label{fd2red}
\end{gather}
as demonstrated in detail in Appendix \ref{app-dyn}. Contracting the above equations with $M^{ab}$ and $\Sigma^{ab}$ gives four $0\!=\!0$ conditions indicating that there is a total of $6\!\times\!2\!-\!4\!=\!8$ field equations. Therefore, compared to the original Dirac equation, they display no redundancy, and the issue is proven to be solved when writing the flag-dipole field equations as in (\ref{fd1red}-\ref{fd2red}).

As for the limiting cases, for the dipoles (\ref{d1}-\ref{d3}) reduce to
\begin{gather}
(B_{\mu}\!\mp\!2P_{\mu})U_{\rho}\varepsilon^{\mu\rho\alpha\nu}\!+\!R^{[\alpha}U^{\nu]}\!=\!0
\end{gather}
while for the flagpoles (\ref{f1}-\ref{f6}) reduce to
\begin{gather}
B_{\pi}U_{\mu}\varepsilon^{\pi\mu\rho\sigma}\!+\!R^{[\rho}U^{\sigma]}\!-\!2mM^{\rho\sigma}\!=\!0
\end{gather}
as is now easy to check.
\subsection{Exact Solutions}\label{sol}
We will now focus only on the flag-dipole spinors in the strict sense, that is excluding the two notable limits.

Our aim, henceforth, is that of finding exact solutions for $\alpha$ as a free field determined uniquely in terms of the field equations (\ref{fd1red}-\ref{fd2red}). Because in this section we are not interested in proving general statements, but only to find some specific solution, we will feel free to make a number of assumptions: these will simplify the search, only at the cost of a reduction of generality that is not relevant for our purposes now. However, for the benefit of readers, we are going to enumerate such assumptions, so that it will be straightforward to see where the reduction of generality lies.

We look for solutions in spherical coordinates, for which the metric is
\begin{gather}
g_{tt}\!=\!1\ \ \ \ \ \ \ \ g_{rr}\!=\!-1\ \ \ \ \ \ \ \ g_{\theta\theta}\!=\!-r^{2}\ \ \ \
\ \ \ \ g_{\varphi\varphi}\!=\!-r^{2}|\!\sin{\theta}|^{2}
\end{gather}
generating the connection
\begin{gather}
\Lambda^{\theta}_{\theta r}\!=\!\Lambda^{\varphi}_{\varphi r}\!=\!1/r\ \ \ \ \ \ \ \
\Lambda^{r}_{\theta\theta}\!=\!-r\ \ \ \ \ \ \ \
\Lambda^{r}_{\varphi\varphi}\!=\!-r|\!\sin{\theta}|^{2}\ \ \ \ \ \ \ \
\Lambda^{\varphi}_{\varphi\theta}\!=\!\cot{\theta}\ \ \ \ \ \ \ \
\Lambda^{\theta}_{\varphi\varphi}\!=\!-\cos{\theta}\sin{\theta}
\end{gather}
having zero curvature. As \emph{first hypothesis}, we take all fields depending only on elevation and radial coordinate.

A \emph{second hypothesis} is to have the velocity parametrized as
\begin{gather}
U_{t}\!=\!e^{a}\ \ \ \ \ \ \ \ \ \ \ \ \ \ \ \ U_{\varphi}\!=\!e^{a}r\sin{\theta}\label{Uvec}:
\end{gather}
this hypothesis corresponds to material motions taking place only along the azimuthal coordinate and being compatible with the $U_{a}U^{a}\!=\!0$ constraint. In it, $a$ is a generic function.

The \emph{third hypothesis} is to choose
\begin{gather}
X_{r}\!=\!-\sin{\gamma}\ \ \ \ \ \ \ \ \ \ \ \ \ \ \ \ X_{\theta}\!=\!r\cos{\gamma}\label{Xvec}
\end{gather}
as the simplest vector $X_{\nu}$ compatible with the $X_{a}X^{a}\!=\!-1$ and $X_{a}U^{a}\!=\!0$ constraints. One more time, $\gamma$ is a generic function: now, conditions $\nabla_{\mu}U_{a}\!=\!R_{ka\mu}U^{k}$ and $\nabla_{\mu}X_{a}\!=\!R_{ka\mu}X^{k}$ can be expanded, once the connection is known, into $32$ relations in which only the tensorial connection is unknown. After that these conditions are inverted, the tensorial connection is found to have $4$ components equal to zero and the others given by
\begin{gather}
R_{t\varphi r}\!=\!r\sin{\theta}\partial_{r}a\ \ \ \ \ \ \ \
\ \ \ \ \ \ \ \ R_{t\varphi\theta}\!=\!r\sin{\theta}\partial_{\theta}a\ \ \ \ \ \ \ \
\ \ \ \ \ \ \ \ R_{r\theta r}\!=\!-r\partial_{r}\gamma\ \ \ \ \ \ \ \
\ \ \ \ \ \ \ \ R_{r\theta\theta}\!=\!-r(\partial_{\theta}\gamma\!+\!1)\label{ten-conn1}
\end{gather}
together with the four groups
\begin{gather}
R_{trt}\!=\!e^{a}\cos{\gamma}\partial_{t}S\ \ \ \ \ \ \ \
\ \ \ \ \ \ \ \ R_{trr}\!=\!e^{a}\cos{\gamma}\partial_{r}S\\
R_{t\theta t}\!=\!e^{a}r\sin{\gamma}\partial_{t}S\ \ \ \ \ \ \ \
\ \ \ \ \ \ \ \ R_{t\theta r}\!=\!e^{a}r\sin{\gamma}\partial_{r}S\\
R_{\varphi rt}\!=\!e^{a}r\sin{\theta}\cos{\gamma}\partial_{t}S\ \ \ \ \ \ \ \
\ \ \ \ \ \ \ \ R_{\varphi rr}\!=\!e^{a}r\sin{\theta}\cos{\gamma}\partial_{r}S\\
R_{\varphi\theta t}\!=\!e^{a}r^{2}\sin{\theta}\sin{\gamma}\partial_{t}S\ \ \ \ \ \ \ \
\ \ \ \ \ \ \ \ R_{\varphi\theta r}\!=\!e^{a}r^{2}\sin{\theta}\sin{\gamma}\partial_{r}S
\end{gather}
\begin{gather}
R_{tr\theta}\!=\!e^{a}\cos{\gamma}\partial_{\theta}S\ \ \ \ \ \ \ \
\ \ \ \ \ \ \ \ \ \ \ \ \ \ \ \ \ \ \ \ R_{tr\varphi}\!=\!e^{a}\cos{\gamma}\partial_{\varphi}S\\
R_{t\theta \theta}\!=\!e^{a}r\sin{\gamma}\partial_{\theta}S\ \ \ \ \ \ \ \
\ \ \ \ \ \ \ \ \ \ \ \ \ \ \ \ \ \ \ \ R_{t\theta\varphi}\!=\!e^{a}r\sin{\gamma}\partial_{\varphi}S\\
R_{\varphi r\theta}\!=\!e^{a}r\sin{\theta}\cos{\gamma}\partial_{\theta}S\ \ \ \ \ \ \ \
\ \ \ \ \ \ \ \ R_{\varphi r\varphi}\!=\!r\sin{\theta}(\sin{\theta}
\!+\!e^{a}\cos{\gamma}\partial_{\varphi}S)\\
R_{\varphi\theta \theta}\!=\!e^{a}r^{2}\sin{\theta}\sin{\gamma}\partial_{\theta}S\ \ \ \ \ \ \ \
\ \ \ \ \ \ \ \ R_{\varphi\theta\varphi}\!=\!r^{2}\sin{\theta}(\cos{\theta}
\!+\!e^{a}\sin{\gamma}\partial_{\varphi}S)
\end{gather}
where $S$ is a generic function.

A \emph{fourth assumption} is to have $S\!=\!0$, so that the tensorial connection reduces to (\ref{ten-conn1}) and
\begin{gather}
R_{r\varphi\varphi}\!=\!-r(\sin{\theta})^{2}\ \ \ \ \ \ \ \ \ \ \ \ \ \ \ \
\ \ \ \ \ \ \ \ R_{\theta\varphi\varphi}\!=\!-r^{2}\sin{\theta}\cos{\theta}\label{ten-conn2}:
\end{gather}
notice that these last two are purely geometric, exactly like the term $-r$ in the $R_{r\theta\theta}$ component of (\ref{ten-conn1}); the remaining terms in all components of (\ref{ten-conn1}) encode instead the derivatives of the $a$ and $\gamma$ functions. From (\ref{ten-conn1}, \ref{ten-conn2}) we get
\begin{gather}
R_{r}=(2+\partial_{\theta}\gamma)/r\ \ \ \ \ \ \ \ \ \ \ \ \ \ \ \
R_{\theta}=\cot{\theta}-r\partial_{r}\gamma\label{R}\\
B_{r}=\partial_{\theta}a/r\ \ \ \ \ \ \ \ \ \ \ \ \ \ \ \
B_{\theta}=-r\partial_{r}a\label{B},
\end{gather}
necessary for the field equations.

As a \emph{fifth assumption}, and the last one, we will take $P_{\nu}\!=\!0$ (this assumption may be unphysical for regular spinors, but interpretations that work for regular spinors can not necessarily be extended to singular spinors: for instance, we have that $P_{\nu}\!\equiv\!0$ identically for flagpoles \cite{Fabbri:2024eec}). With this assumption, and using (\ref{M}), equations (\ref{fd1red}-\ref{fd2red}) become
\begin{gather}
B_{\pi}U_{\mu}\varepsilon^{\pi\mu\rho\sigma}
\!+\!R^{[\rho}U^{\sigma]}
\!-\!2m\cos{\alpha}U^{[\rho}X^{\sigma]}\!=\!0\\
U_{[\rho}\nabla_{\sigma]}\alpha
\!-\!2m\sin{\alpha}U_{[\rho}X_{\sigma]}\!=\!0,
\end{gather}
so that now, plugging (\ref{R}-\ref{B}), they can explicitly be written down, component-by-component: after this is done, four equations are identically verified, and the remaining $8$ are given by
\begin{gather}
\partial_{r}\alpha\!+\!2m\sin{\gamma}\sin{\alpha}\!=\!0
\label{FE1}\\
\partial_{\theta}\alpha\!-\!2mr\cos{\gamma}\sin{\alpha}\!=\!0
\label{FE2}\\
\partial_{r}a\!+\!2/r\!+\!\partial_{\theta}\gamma/r\!-\!2m\sin{\gamma}\cos{\alpha}\!=\!0
\label{FE3}\\
\partial_{\theta}a\!+\!\cot{\theta}\!-\!r\partial_{r}\gamma\!+\!2mr\cos{\gamma}\cos{\alpha}\!=\!0
\label{FE4}
\end{gather}
as is clear after a very straightforward substitution.

As discussed in Appendix \ref{app-int}, the above field equations are solvable only when either $\gamma\!=\!\pm\pi/2$ or $\gamma\!=\!-\theta$ hold:

In the first case, $\gamma\!=\!-\pi/2$ makes (\ref{FE1}-\ref{FE4}) become
\begin{gather}
\partial_{r}\alpha\!-\!2m\sin{\alpha}\!=\!0\\
\partial_{\theta}\alpha\!=\!0\\
\partial_{r}a\!+\!2/r\!+\!2m\cos{\alpha}\!=\!0\\
\partial_{\theta}a\!+\!\cot{\theta}\!=\!0:
\end{gather}
they are solved by
\begin{gather}
e^{a}=\frac{\cosh{(2mr)}}{(2mr)^{2}\sin{\theta}}
\end{gather}
and
\begin{gather}
\sin{\alpha}=\frac{1}{\cosh{(2mr)}}\ \ \ \ \ \ \ \ \ \ \ \ \ \ \ \
\cos{\alpha}=-\tanh{(2mr)}\label{solution-sph}
\end{gather}
(the case $\gamma\!=\!\pi/2$ would have given analogous results).

In the second case, $\gamma\!=\!-\theta$ makes (\ref{FE1}-\ref{FE4}) reduce to
\begin{gather}
\partial_{r}a\!+\!1/r\!+\!2m\sin{\theta}\cos{\alpha}\!=\!0\\
\partial_{r}\alpha\!-\!2m\sin{\theta}\sin{\alpha}\!=\!0\\
\partial_{\theta}a\!+\!\cot{\theta}\!+\!2mr\cos{\theta}\cos{\alpha}\!=\!0\\
\partial_{\theta}\alpha\!-\!2mr\cos{\theta}\sin{\alpha}\!=\!0:
\end{gather}
these are solved by
\begin{gather}
e^{a}=\frac{\cosh{(2mr\sin{\theta})}}{2mr\sin{\theta}}
\end{gather}
and
\begin{gather}
\sin{\alpha}=\frac{1}{\cosh{(2mr\sin{\theta})}}\ \ \ \ \ \ \ \ \ \ \ \ \ \ \ \
\cos{\alpha}=-\tanh{(2mr\sin{\theta})}\label{solution-cyl}
\end{gather}
as straightforward to check.

The physical field $\alpha$ has spherical symmetry in the former case, but only cylindrical symmetry in the latter case.

Interested readers might find it helpful to have this solution written also in standard form. For that, we begin with the remark that in the frame where $\boldsymbol{L}\!=\!\mathbb{I}$ we have $X^{2}\!=\!1$ and $U_{0}\!=\!U_{3}\!=\!1$ (see Appendix \ref{app-M}). In this reference frame, the soldering between these expressions for $X_{a}$ and $U_{a}$ (in Latin, Lorentz indices) and (\ref{Xvec})-(\ref{Uvec}) for $X_{\nu}$ and $U_{\nu}$ (with Greek, coordinate indices) is given, for the case $\gamma\!=\!-\pi/2$, by the tetrads fields
\begin{eqnarray}
&\xi^{0}_{t}\!=\!\cosh{a}\ \ \ \ \ \ \ \ \xi^{3}_{t}\!=\!\sinh{a}\ \ \ \
\ \ \ \ \xi_{0}^{t}\!=\!\cosh{a}\ \ \ \ \ \ \ \ \xi_{3}^{t}\!=\!-\sinh{a}\\
&\xi^{2}_{r}\!=\!-1\ \ \ \
\ \ \ \ \xi_{2}^{r}\!=\!-1\\
&\xi^{1}_{\theta}\!=\!r\ \ \ \
\ \ \ \ \ \ \ \ \xi_{1}^{\theta}\!=\!\frac{1}{r}\\
&\xi^{0}_{\varphi}\!=\!r\sin{\theta}\sinh{a}\ \ \ \ \ \ \ \ \xi^{3}_{\varphi}\!=\!r\sin{\theta}\cosh{a}\ \ \ \ \ \ \ \ \xi_{0}^{\varphi}\!=\!-\frac{1}{r\sin{\theta}}\sinh{a}\ \ \ \ \ \ \ \ \xi_{3}^{\varphi}\!=\!\frac{1}{r\sin{\theta}}\cosh{a}
\end{eqnarray}
generating the spin-connection
\begin{gather}
C_{03r}\!=\!-\partial_{r}a\ \ \ \ \ \ \ \ \ \ \ \ C_{03\theta}\!=\!-\partial_{\theta}a\ \ \ \ \ \ \ \ \ \ \ \ C_{21\theta}\!=\!-1\label{BG-sph1}\\
C_{02\varphi}\!=\!-\sin{\theta}\sinh{a}\ \ \ \ \ \ \ \
C_{01\varphi}\!=\!\cos{\theta}\sinh{a}\ \ \ \ \ \ \ \
C_{31\varphi}\!=\!-\cos{\theta}\cosh{a}\ \ \ \ \ \ \ \
C_{23\varphi}\!=\!-\sin{\theta}\cosh{a}\label{BG-sph2}
\end{gather}
of course with zero curvature.

The soldering is given instead, for the case $\gamma\!=\!-\theta$, by the tetrads fields
\begin{eqnarray}
&\xi^{0}_{t}\!=\!\cosh{a}\ \ \ \ \ \ \ \ \xi^{3}_{t}\!=\!\sinh{a}\ \ \ \
\ \ \ \ \xi_{0}^{t}\!=\!\cosh{a}\ \ \ \ \ \ \ \ \xi_{3}^{t}\!=\!-\sinh{a}\\
&\xi^{2}_{r}\!=\!-\sin{\theta}\ \ \ \ \ \ \ \ \xi^{1}_{r}\!=\!-\cos{\theta}\ \ \ \
\ \ \ \ \xi_{2}^{r}\!=\!-\sin{\theta}\ \ \ \ \ \ \ \ \xi_{1}^{r}\!=\!-\cos{\theta}\\
&\xi^{2}_{\theta}\!=\!-r\cos{\theta}\ \ \ \ \ \ \ \ \xi^{1}_{\theta}\!=\!r\sin{\theta}\ \ \ \
\ \ \ \ \xi_{2}^{\theta}\!=\!-\frac{1}{r}\cos{\theta}\ \ \ \ \ \ \ \ \xi_{1}^{\theta}\!=\!\frac{1}{r}\sin{\theta}\\
&\xi^{0}_{\varphi}\!=\!r\sin{\theta}\sinh{a}\ \ \ \ \ \ \ \ \xi^{3}_{\varphi}\!=\!r\sin{\theta}\cosh{a}\ \ \ \ \ \ \ \ \xi_{0}^{\varphi}\!=\!-\frac{1}{r\sin{\theta}}\sinh{a}\ \ \ \ \ \ \ \ \xi_{3}^{\varphi}\!=\!\frac{1}{r\sin{\theta}}\cosh{a}
\end{eqnarray}
generating the spin-connection
\begin{gather}
C_{03r}\!=\!-\partial_{r}a\ \ \ \ \ \ \ \ C_{03\theta}\!=\!-\partial_{\theta}a\ \ \ \ \ \ \ \
C_{02\varphi}\!=\!-\sinh{a}\ \ \ \ \ \ \ \ C_{23\varphi}\!=\!-\cosh{a}\label{BG-cyl}
\end{gather}
again with zero curvature.

In the background of the spin-connection (\ref{BG-sph1}-\ref{BG-sph2}), the Dirac equation (\ref{D}) acquires the structure
\begin{eqnarray}
&-2i\boldsymbol{\gamma}^{2}\left(r\partial_{r}\psi\!+\!\psi\right)
\!+\!i\cot{\theta}\boldsymbol{\gamma}^{1}\psi
\!+\!r\partial_{r}a\boldsymbol{\pi}\boldsymbol{\gamma}^{1}\psi
\!+\!\partial_{\theta}a\boldsymbol{\pi}\boldsymbol{\gamma}^{2}\psi
\!-\!2mr\psi=0\label{Dexpl-sph}.
\end{eqnarray}

In the background of the spin-connection (\ref{BG-cyl}), the Dirac equation (\ref{D}) is instead
\begin{eqnarray}
\nonumber
&i\boldsymbol{\gamma}^{1}\left(2\sin{\theta}\partial_{\theta}\psi
\!-\!2\cos{\theta}r\partial_{r}\psi\right)
\!-\!i\boldsymbol{\gamma}^{2}\left(2\cos{\theta}\partial_{\theta}\psi
\!+\!2\sin{\theta}r\partial_{r}\psi\!+\!\csc{\theta}\psi\right)+\\
&+\left(\partial_{\theta}a\sin{\theta}\!-\!r\partial_{r}a\cos{\theta}\right)
\boldsymbol{\pi}\boldsymbol{\gamma}^{2}\psi
\!+\!\left(\partial_{\theta}a\cos{\theta}\!+\!r\partial_{r}a\sin{\theta}\right)
\boldsymbol{\pi}\boldsymbol{\gamma}^{1}\psi\!-\!2mr\psi=0\label{Dexpl-cyl}.
\end{eqnarray}

The first solution (\ref{solution-sph}) in its standard form corresponds to the spinor field
\begin{gather}
\psi\!=\!\frac{1}{\sqrt{\cosh{(2mr)}}}\left(\begin{tabular}{c}
$\cosh{(mr)}$\\
$0$\\
$0$\\
$-\sinh{(mr)}$
\end{tabular}\right)\label{spinorsol-sph},
\end{gather}
in terms of which (\ref{Dexpl-sph}) results into
\begin{gather}
\left[2mr\left(\tanh{(2mr)}\mathbb{I}\!-\!\boldsymbol{\pi}\boldsymbol{\gamma}^{1}\right)
\!-\!2\mathbb{I}\!+\!i\cot{\theta}\boldsymbol{\pi}\right](i\boldsymbol{\gamma}^{2}\!+\!\boldsymbol{\pi}\boldsymbol{\gamma}^{1})\psi\!=\!0
\end{gather}
which is clearly satisfied.

The second solution (\ref{solution-cyl}) in standard form corresponds to the spinor field given by
\begin{gather}
\psi\!=\!\frac{1}{\sqrt{\cosh{(2mr\sin{\theta})}}}\left(\begin{tabular}{c}
$\cosh{(mr\sin{\theta})}$\\
$0$\\
$0$\\
$-\sinh{(mr\sin{\theta})}$
\end{tabular}\right)\label{spinorsol-cyl},
\end{gather}
in terms of which (\ref{Dexpl-cyl}) results into
\begin{gather}
\left[\left(\tanh{(2mr\sin{\theta})}\!-\!\frac{1}{2mr\sin{\theta}}\right)\mathbb{I}\!-\!\boldsymbol{\pi}\boldsymbol{\gamma}^{1}\right](i\boldsymbol{\gamma}^{2}\!+\!\boldsymbol{\pi}\boldsymbol{\gamma}^{1})\psi\!=\!0
\end{gather}
which is clearly verified.

The two solutions (\ref{spinorsol-sph}) and (\ref{spinorsol-cyl}) have the same structure
\begin{gather}
\psi\!=\!\frac{1}{\sqrt{\cosh{(2x)}}}\left(\begin{tabular}{c}
$\cosh{x}$\\
$0$\\
$0$\\
$-\sinh{x}$
\end{tabular}\right)\label{spinorsol}
\end{gather}
with $x\!=\!mr$ in the first case (spherical symmetry) and $x\!=\!mr\sin{\theta}$ in the second case (cylindrical symmetry).

They are exact solutions of the Dirac equation, but because they give rise to $\Phi\!=\!\Theta\!=\!0$ identically then their spinor is not a Dirac field. On the other hand, we have
\begin{eqnarray}
(M^{ab})\!=\!\tanh{(2x)}\left(\begin{array}{cccc}
0 & 0 & -1 & 0\\
0 & 0 & 0 & 0\\
1 & 0 & 0 & -1\\
0 & 0 & 1 & 0
\end{array}\right)\!\neq\!0,
\end{eqnarray}
so the spinor is also not Weyl, and
\begin{eqnarray}
(S^{a})\!=\!\frac{1}{\cosh{(2x)}}\left(\begin{array}{c}
-1\\
0\\
0\\
1
\end{array}\right)\!\neq\!0,
\end{eqnarray}
so the spinor is also not Majorana. More in detail, the solutions are not Dirac since the two chiral parts are orthogonal (that is, they have opposite helicities). However, the two chiral parts have different modules ($\cosh{x}$ for the left and $\sinh{x}$ for the right), so that they are neither Weyl (for which one of the two modules is zero) nor Majorana (for which the two modules are equal). In other words, the above solutions cannot be Weyl since they possess both chiral parts but they also cannot be Majorana since the two chiral parts are not the conjugated of one another.

Whether or not in cylindrical coordinates, the third direction is singled out as privileged also because of the presence of the third component of the spin $S^{3}\!=\!1/\cosh{(2x)}$; of the angular momentum, the ``electric'' part $M^{02}\!=\!-\tanh{(2x)}$ is directed along the second axis while the ``magnetic'' part $M^{23}\!=\!-\tanh{(2x)}$ is directed along the first axis. Remark that the spin has the behaviour of a hyperbolic secant (a soliton), while the angular momentum behaves as a hyperbolic tangent (topological soliton). This behaviour might be interesting when interacting with electrodynamics.

As a last comment, it may be of interest to point out that, for this type of flag-dipole spinor, the tensorial connection $R_{ab\nu}$ can never be equal to zero. In fact, by setting to zero all $24$ components, it is straightforward to see that not all the resulting $24$ conditions can be verified at once. For flag-dipoles, non-zero tensorial connections are necessary.

More details about these two solutions may come from a general analysis on stability (presumably, the methods used in dynamical systems might have something to say about this aspect). Possible extensions might come from the interaction with abelian gauge fields at least in the cases in which condition $P_{i}\!\neq\!0$ is allowed.

The solution we found is a special solution restricted by five hypotheses: 1. static cylindrical symmetry, 2. velocity parametrized as (\ref{Uvec}), 3. auxiliary vector $X_{i}$ chosen as (\ref{Xvec}), 4. tensorial connection as (\ref{ten-conn1}) and (\ref{ten-conn2}), 5. $P_{i}\!=\!0$ --- any of these five assumptions can be dropped to enlarge the space of solutions. Although in the present paper we needed to find only some solution as proof of concept, the quest for more general solutions is an open one.
\section{Conclusion}
In this work, we have considered the Dirac equation, finding exact solutions that are neither Dirac, nor Weyl nor Majorana spinors: unlike Weyl, they display both chiralities, and unlike Majorana, the two chiralities are independent from each other; but unlike Dirac, the two chiralities have opposite helicities. The found spinor is a flag-dipole, whose spin axial-vector is still aligned along the third axis, but whose angular momentum has electric and magnetic parts that are orthogonal to each other and to the spin axial-vector. The spin axial-vector itself was found to have the field distribution of a hyperbolic secant while the angular momentum was found to be a hyperbolic tangent.

These solutions were two, one displaying spherical symmetry and the other with cylindrical symmetry. Both however were obtained by insisting on five restrictions (no time or azimuthal dependence, velocity parametrized as in (\ref{Uvec}), $X_{a}$ vector as in (\ref{Xvec}), trivially-satisfied integrability of the tensorial connection, and absence of linear momentum). Hence the two solutions are quite restricted, and it should be easy to enlarge the underlying background. For more general tensorial connections, for example, or for non-zero linear momentum, computations should still be manageable.

Can the above solutions, or any generalization, be realized in labs? And if no flag-dipole is physical, then how is it that the two limiting cases (Weyl and Majorana spinors) appears to be?
\vspace{10pt}

\textbf{Funding}. This work is funded by Next Generation EU via the project ``Geometrical and Topological effects on Quantum Matter (GeTOnQuaM)'' in the framework of activities of the INFN Research Project QGSKY.

\

\textbf{Data Availability}. The manuscript does not have associated data in any repository.

\

\textbf{Conflict of interest}. There is no conflict of interest.
\appendix
\section{Expression of the angular momentum}\label{app-M}
In this first appendix, we prove expression (\ref{M}). To do that, first consider the fact that (\ref{M}) is a tensorial expression, and as such it is generally valid if it is proved to be valid in one system of reference. For this purpose, it is enough to prove that it is valid in the system of reference in which $\boldsymbol{L}\!=\!\mathbb{I}$, where
\begin{gather}
M^{02}\!=\!M^{23}\!=\!\cos{\alpha}\\
U_{0}\!=\!U_{3}\!=\!1
\end{gather}
as shown in \cite{Fabbri:2020elt}. In this frame, expression (\ref{M}) is written component-by-component as
\begin{gather}
0\!=\!\cos{\alpha}X^{1}\ \ \ \ \ \ \ \ 0\!=\!0\\
1\!=\!X^{2}\ \ \ \ \ \ \ \ 0\!=\!\cos{\alpha}X^{1}\\
0\!=\!\cos{\alpha}(X^{3}\!+\!X^{0})\ \ \ \ \ \ \ \ 1\!=\!X^{2}
\end{gather}
which can be solved, for example, by the vector
\begin{gather}
X^{0}\!=\!0\ \ \ \ \ \ \ \ X^{1}\!=\!0\ \ \ \ \ \ \ \ X^{2}\!=\!1\ \ \ \ \ \ \ \ X^{3}\!=\!0
\end{gather}
as easy to see. Clearly, we also have that $X_{a}X^{a}\!=\!-1$, as well as $X_{a}U^{a}\!=\!0$, in the same frame. Therefore, we have that this expression of $X_{a}$ makes (\ref{M}) is valid, in the frame where $\boldsymbol{L}\!=\!\mathbb{I}$ and therefore, due to its covariant character, (\ref{M}) is valid in general (the expression of $X_{a}$ in generic frames is then obtained by applying a real Lorentz transformation $\Lambda^{a}_{b}$ given by $\boldsymbol{L}\boldsymbol{\gamma}^{b}\boldsymbol{L}^{-1}\Lambda^{a}_{b}\!=\!\boldsymbol{\gamma}^{a}$ as the usual passage from complex to real representations).
\section{Identically-verified derivative of the angular momentum}\label{app-der}
Let us next consider $\nabla_{\mu}X_{a}\!=\!R_{ka\mu}X^{k}$ and prove, together with (\ref{dU}), as well as (\ref{M}), that (\ref{dM}) is true.

The proof is a mere computation, starting from (\ref{dM}) written as
\begin{gather}
\nabla_{\mu}M^{ab}\!+\!M^{ab}\tan{\alpha}\nabla_{\mu}\alpha
\!+\!R^{a}_{\phantom{a}k\mu}M^{kb}\!-\!R^{b}_{\phantom{b}k\mu}M^{ka}\!=\!0
\end{gather}
and verifying its vanishing. We have in fact that substituting (\ref{M}) gives
\begin{gather}
\nonumber
\nabla_{\mu}M^{ab}\!+\!M^{ab}\tan{\alpha}\nabla_{\mu}\alpha
\!+\!R^{a}_{\phantom{a}k\mu}M^{kb}\!-\!R^{b}_{\phantom{b}k\mu}M^{ka}=\\
\nonumber
=\nabla_{\mu}(\cos{\alpha}U^{[a}X^{b]})\!+\!\cos{\alpha}U^{[a}X^{b]}\tan{\alpha}\nabla_{\mu}\alpha
\!+\!R^{a}_{\phantom{a}k\mu}\cos{\alpha}U^{[k}X^{b]}\!-\!R^{b}_{\phantom{b}k\mu}\cos{\alpha}U^{[k}X^{a]}=\\
\nonumber
=-\sin{\alpha}\nabla_{\mu}\alpha U^{[a}X^{b]}
\!+\!\cos{\alpha}\nabla_{\mu}U^{a}X^{b}
\!+\!\cos{\alpha}U^{a}\nabla_{\mu}X^{b}
\!-\!\cos{\alpha}\nabla_{\mu}U^{b}X^{a}
\!-\!\cos{\alpha}U^{b}\nabla_{\mu}X^{a}+\\
\nonumber
+\sin{\alpha}\nabla_{\mu}\alpha U^{[a}X^{b]}
\!-\!\cos{\alpha}R^{ka}_{\phantom{ka}\mu}U_{k}X^{b}
\!+\!\cos{\alpha}R^{ka}_{\phantom{ka}\mu}U^{b}X_{k}
\!+\!\cos{\alpha}R^{kb}_{\phantom{kb}\mu}U_{k}X^{a}
\!-\!\cos{\alpha}R^{kb}_{\phantom{kb}\mu}U^{a}X_{k}:
\end{gather}
plugging next $\nabla_{\mu}X_{a}\!=\!R_{ka\mu}X^{k}$ and (\ref{dU}), and simplifying, we find that the last two lines cancel, leaving zero.
\section{Field equation reduction and their redundancy}\label{app-dyn}
We now prove the fact that field equations (\ref{fd1}-\ref{fd4}) can be reduced to field equations (\ref{fd1red}-\ref{fd2red}). For this, start from the Dirac field equation (\ref{D}), with covariant derivative written in terms of (\ref{decspinder-sing}), as
\begin{gather}
\left[\left(i\tan{\alpha}\nabla_{\mu}\alpha\!-\!iR_{\mu}\!-\!2P_{\mu}\right)
\boldsymbol{\gamma}^{\mu}
\!+\!\left(i\sec{\alpha}\nabla_{\mu}\alpha\!-\!B_{\mu}\right)
\boldsymbol{\gamma}^{\mu}\boldsymbol{\pi}
\!+\!2m\mathbb{I}\right]\psi\!=\!0\label{Dapp}
\end{gather}
in which identity $\boldsymbol{\gamma}_{a}\boldsymbol{\gamma}_{b}\boldsymbol{\gamma}_{c}\!=\!
\boldsymbol{\gamma}_{a}\eta_{bc}\!-\!\boldsymbol{\gamma}_{b}\eta_{ac}\!+\!\boldsymbol{\gamma}_{c}\eta_{ab}\!-\!i\varepsilon_{abci}\boldsymbol{\gamma}^{i}\boldsymbol{\pi}$ was used. Contracting next with $2i\overline{\psi}\boldsymbol{\sigma}^{\rho\sigma}$ on the left, then splitting real and imaginary parts, provides
\begin{gather}
(2\sin{\alpha}P_{\pi}\!-\!B_{\pi})U_{\mu}\varepsilon^{\pi\mu\rho\sigma}
\!-\!R^{[\rho}U^{\sigma]}
\!+\!2mM^{\rho\sigma}\!=\!0\label{app1}\\
(2P\!-\!\sin{\alpha}B)^{[\rho}U^{\sigma]}
\!+\!(\cos{\alpha}\nabla_{\pi}\alpha\!+\!\sin{\alpha}R_{\pi})
U_{\mu}\varepsilon^{\pi\mu\rho\sigma}\!=\!0\label{app2}
\end{gather}
where now identity $2\boldsymbol{\sigma}_{\rho\sigma}\boldsymbol{\gamma}_{\mu}\!=\!
\boldsymbol{\gamma}_{\rho}g_{\sigma\mu}\!-\!\boldsymbol{\gamma}_{\sigma}g_{\rho\mu}\!-\!i\varepsilon_{\rho\sigma\mu\pi}\boldsymbol{\gamma}^{\pi}\boldsymbol{\pi}$ was used. Combining (\ref{app1}) with the Hodge dual of (\ref{app2}) gives
\begin{gather}
U_{[\rho}\nabla_{\sigma]}\alpha
\!+\!2\cos{\alpha}P^{\pi}U^{\mu}\varepsilon_{\pi\mu\rho\sigma}
\!-\!2m\tan{\alpha}M_{\rho\sigma}\!=\!0\label{app3}:
\end{gather}
expressions (\ref{app1}) and (\ref{app3}) are exactly (\ref{fd1red}-\ref{fd2red}). Take them now both in their initial form
\begin{gather}
(2\sin{\alpha}P_{\pi}\!-\!B_{\pi})U_{\mu}\varepsilon^{\pi\mu\rho\sigma}
\!-\!R^{[\rho}U^{\sigma]}
\!+\!2mM^{\rho\sigma}\!=\!0\\
U_{[\rho}\nabla_{\sigma]}\alpha
\!+\!2\cos{\alpha}P^{\pi}U^{\mu}\varepsilon_{\pi\mu\rho\sigma}
\!-\!2m\tan{\alpha}M_{\rho\sigma}\!=\!0
\end{gather}
and in their Hodge duals
\begin{gather}
(2\sin{\alpha}P\!-\!B)_{[\eta}U_{\zeta]}
\!+\!R^{\rho}U^{\sigma}\varepsilon_{\rho\sigma\eta\zeta}
\!-\!mM^{\rho\sigma}\varepsilon_{\rho\sigma\eta\zeta}\!=\!0\\
U_{\rho}\nabla_{\sigma}\alpha\varepsilon^{\rho\sigma\eta\zeta}
\!-\!2\cos{\alpha}P^{[\eta}U^{\zeta]}
\!-\!m\tan{\alpha}M_{\rho\sigma}\varepsilon^{\rho\sigma\eta\zeta}\!=\!0:
\end{gather}
contracting these four on $U_{i}$ one gets
\begin{gather}
R_{\rho}U^{\rho}\!=\!0\label{app4}\\
U^{\rho}\nabla_{\rho}\alpha\!=\!0\label{app5}\\
B_{\eta}U^{\eta}\!=\!0\label{app6}\\
P_{\eta}U^{\eta}\!=\!0\label{app7}
\end{gather}
while contracting them on $X_{i}$ one gets
\begin{gather}
(2\sin{\alpha}P\!-\!B)_{\pi}U_{\mu}X_{\sigma}\varepsilon^{\rho\pi\mu\sigma}
\!+\!U^{\rho}X^{\sigma}R_{\sigma}
\!-\!2m\cos{\alpha}U^{\rho}\!=\!0\\
U_{\rho}X^{\sigma}\nabla_{\sigma}\alpha
\!+\!2\cos{\alpha}P^{\pi}U^{\mu}X^{\sigma}\varepsilon_{\rho\pi\mu\sigma}
\!+\!2m\sin{\alpha}U_{\rho}\!=\!0\\
(2\sin{\alpha}P\!-\!B)^{\eta}X_{\eta}U_{\zeta}
\!+\!R^{\rho}U^{\sigma}X^{\eta}\varepsilon_{\rho\sigma\eta\zeta}\!=\!0\\
U^{\rho}X^{\eta}\nabla^{\sigma}\alpha\varepsilon_{\zeta\rho\eta\sigma}
\!-\!2\cos{\alpha}P^{\eta}X_{\eta}U_{\zeta}\!=\!0
\end{gather}
after using (\ref{Zero}) as well as (\ref{U}). Finally, employing (\ref{app4}-\ref{app7}), the last four become
\begin{gather}
(2\sin{\alpha}P\!-\!B)_{\mu}\Sigma^{\mu\nu}
\!-\!R_{\mu}M^{\mu\nu}
\!-\!2m(\cos{\alpha})^{2}U^{\nu}\!=\!0\\
-M^{\mu\nu}\nabla_{\mu}\alpha\sec{\alpha}
\!+\!2P_{\mu}\Sigma^{\mu\nu}
\!+\!2m\sin{\alpha}U^{\nu}\!=\!0\\
(2\sin{\alpha}P\!-\!B)_{\mu}M^{\mu\nu}
\!+\!R_{\mu}\Sigma^{\mu\nu}\!=\!0\\
\Sigma^{\mu\nu}\nabla_{\mu}\alpha\sec{\alpha}
\!+\!2P_{\mu}M^{\mu\nu}\!=\!0
\end{gather}
in which (\ref{Sigma}) and (\ref{M}) were used. These can be simply worked out to reach
\begin{gather}
B_{\mu}\Sigma^{\mu\nu}
\!+\!R_{\mu}M^{\mu\nu}
\!-\!M^{\mu\nu}\nabla_{\mu}\alpha\tan{\alpha}
\!+\!2mU^{\nu}\!=\!0\\
-M^{\mu\nu}\nabla_{\mu}\alpha\sec{\alpha}
\!+\!2P_{\mu}\Sigma^{\mu\nu}
\!+\!2m\sin{\alpha}U^{\nu}\!=\!0\\
-B_{\mu}M^{\mu\nu}
\!+\!R_{\mu}\Sigma^{\mu\nu}
\!-\!\Sigma^{\mu\nu}\nabla_{\mu}\alpha\tan{\alpha}\!=\!0\\
\Sigma^{\mu\nu}\nabla_{\mu}\alpha\sec{\alpha}
\!+\!2P_{\mu}M^{\mu\nu}\!=\!0
\end{gather}
which are just (\ref{fd1}-\ref{fd4}), albeit in a shuffled ordering.

And as was proven in \cite{Fabbri:2020elt}, (\ref{fd1}-\ref{fd4}) imply back the original Dirac equation (\ref{D}).

Consequently, the Dirac equation (\ref{D}), the polar form (\ref{fd1}-\ref{fd4}) and the polar form (\ref{fd1red}-\ref{fd2red}) are equivalent.
\section{Integrability conditions of the field equations}\label{app-int}
Field equations (\ref{FE1}-\ref{FE4}) can be worked out to provide some conditions for integrability. We begin by noticing that, from (\ref{FE1}) and (\ref{FE2}) written as
\begin{gather}
\partial_{r}\alpha\!=\!-2m\sin{\gamma}\sin{\alpha}\\
\partial_{\theta}\alpha\!=\!2mr\cos{\gamma}\sin{\alpha},
\end{gather}
the integrability conditions $\partial_{\theta}\partial_{r}\alpha\!=\!\partial_{r}\partial_{\theta}\alpha$ give immediately
\begin{gather}
(\partial_{\theta}\gamma\!+\!1)\cos{\gamma}\!=\!r\partial_{r}\gamma\sin{\gamma}\label{aux}.
\end{gather}

Then, from (\ref{FE3}) and (\ref{FE4}), the integrability condition $\partial_{\theta}\partial_{r}a\!=\!\partial_{r}\partial_{\theta}a$ gives
\begin{gather}
\partial_{r}(r\partial_{r}\gamma)\!+\!\partial_{\theta}\partial_{\theta}\gamma/r
\!=\!2m\cos{\alpha}(\cos{\gamma}\partial_{\theta}\gamma
\!+\!\cos{\gamma}\!-\!r\sin{\gamma}\partial_{r}\gamma)
\end{gather}
after (\ref{FE1}) and (\ref{FE2}) have been plugged in. Then employing (\ref{aux}) allows us to reach
\begin{gather}
r\partial_{r}(r\partial_{r}\gamma)\!+\!\partial_{\theta}\partial_{\theta}\gamma\!=\!0
\end{gather}
spelling that the function $\gamma$ is harmonic.

The only harmonic functions that also verify (\ref{aux}) are the functions of $\theta$ for which
\begin{gather}
(\partial_{\theta}\gamma\!+\!1)\cos{\gamma}\!=\!0
\end{gather}
and therefore $\gamma\!=\!-\theta$ and $\gamma\!=\!\pm\pi/2$ are the only choices.

\end{document}